\documentclass[12pt]{article}

\usepackage{epsfig}

\usepackage{amssymb}
\usepackage{amsfonts}

\usepackage{color}
 
\def\be{\begin{equation}}
\def\ee{\end{equation}}
\def\ba{\begin{array}{c}}
\def\ea{\end{array}}

\newcommand{\bea}{\begin{eqnarray}}
\newcommand{\eea}{\end{eqnarray}}

\newcommand{\kt}{\rangle}

\begin{document}

\begin{center}

.

{\Large \bf

Quantum mechanics using two auxiliary inner products

}

\vspace{20mm}

 {\bf Miloslav Znojil}

 \vspace{3mm}
Nuclear Physics Institute ASCR, Hlavn\'{\i} 130, 250 68 \v{R}e\v{z},
Czech Republic

\vspace{0.2cm}

 and

\vspace{0.2cm}

Department of Physics, Faculty of Science, University of Hradec
Kr\'{a}lov\'{e},

Rokitansk\'{e}ho 62, 50003 Hradec Kr\'{a}lov\'{e},
 Czech Republic

{e-mail: znojil@ujf.cas.cz}


\newpage


\end{center}

\subsection*{Abstract}

The current applications of
non-Hermitian but
${\cal PT}-$symmetric Hamiltonians $H$
cover several, mutually not too
closely connected subdomains of
quantum physics.
Mathematically, the
split between the open and closed systems
can be characterized
by the respective triviality and non-triviality of
an auxiliary inner-product
metric $\Theta=\Theta(H)$.
With our attention restricted to
the latter, mathematically more interesting unitary-evolution case
we show that the intuitive but technically decisive
simplification of the theory
achieved via an ``additional'' ${\cal PCT}-$symmetry
constraint upon $H$ can be given
a deeper mathematical meaning via introduction
of a certain second auxiliary inner product.

\subsection*{Keywords}

.

formulations of quantum mechanics;

Schr\"{o}dinger representation;

non-Hermitian Hamiltonians;

${\cal PT}-$ and ${\cal PCT}-$symmetry;

physical and auxiliary inner products;

\newpage

\section{Introduction}

In the textbooks on quantum mechanics
the basic features of bound states $|\psi\kt$
are best illustrated
via ordinary differential Hamiltonian operators
  \be
  H = -\frac{d^2}{dx^2} +V(x)\,
  \,
 \label{BGr}
 \ee
containing various
real and confining one-dimensional potentials $V(x)$ \cite{Messiah}.
Operators (\ref{BGr}) are assumed
acting in the most common physical
Hilbert space $L^2(\mathbb{R})$
of complex square-integrable functions.
In 1998, Bender with Boettcher \cite{BB}
attracted attention of the physics community to certain
less conventional
Hamiltonians (\ref{BGr})
in which the potential was complex.
The Hamiltonian
became non-Hermitian
but its spectrum remained
real, discrete and bounded from below, i.e., compatible with the
possible unitarity of the evolution of the
system in question,
in principle at least.
After a replacement of the requirements of
Hermiticity by the intuitively more acceptable
conditions of
${\cal PT}-$symmetry
 \be
 H\,{\cal PT}  = {\cal PT}\,H
 \label{pts}
 \ee
plus ${\cal PCT}-$symmetry
 \be
 H\,{\cal PCT}  = {\cal PCT}\,H
 \label{pcts}
 \ee
(where ${\cal P}$ is parity and ${\cal C}$ is charge while
${\cal T}$ stands for the time reversal)
the unusual choice of the specific non-Hermitian Hamiltonians
has very soon been shown fully
compatible with the
unitarity of dynamics.
A new, antilinear-symmetry-based formulation of quantum mechanics
in Schr\"{o}dinger representation
has been born \cite{Carl,ali,book}.

Currently, the
family of innovative non-Hermitian
models
is increasingly popular
in several branches of physics
\cite{Christodoulides,Carlbook}.
In our present letter we intend to support
this trend by a comment
on some interesting mathematical inner-product structures
emerging behind the new paradigm.
For introduction we have to
add
a brief but important remark on
the terminology. The point
(or rather an unexpected difficulty) is
that even during the very short history of the
new paradigm, its scope and range were already split into
two different
sub-paradigms. In the literature, unfortunately,
both of them
share the same name of
${\cal PT}-$symmetric quantum mechanics.
As long as
such a terminological confluence
became a source of multiple
misunderstandings,
we feel urged to
avoid the potential confusion by distinguishing,
strictly, between the open-system
${\cal PT}-$symmetric quantum
theory (OST) and the closed-system
${\cal PT}-$symmetric quantum
theory (CST).
Exclusively, our attention will be paid to
the latter approach.

Another, closely connected introductory remark
should be added emphasizing the
deep
difference between the physics, background, motivation and impact
of the respective OST and CST studies.
Indeed, the most characteristic phenomenological
characteristics of the OST approach lies in its interest
in the description of various
non-unitary forms of quantum evolution
covering the
resonant and/or dissipative processes.
In some sense, the OST approach is both
mathematically more straightforward
(working just with a unique, single inner product)
and, phenomenologically,
much more traditional
(with one of its roots being the Feshbach's effective,
model-subspace description
of systems living in a larger Hilbert space).
In contrast,
the CST approach can be perceived as
a much younger part of quantum physics
(according to review \cite{MZbook},
hardly taken too seriously before the
publication of the pioneering letter \cite{BB} in 1998).
At the same time, the CST formalism
is currently a true challenge even for
mathematicians (the mathematics-oriented book \cite{book}
can be recalled for the first reading).
For both of these reasons,
our present letter will exclusively be devoted to the
latter, CST theory.
Our marginal supportive argument is that
only this form of the theory is fundamental,
dealing with a complete information about the dynamics,
and concerning just the
description of the
unitarily evolving quantum systems.

A brief remark should be finally added emphasizing the innovative
aspects of the CST-related physics. The continuous emergence of open
questions offered also a basic motivation of our present study. Two
of its aspects have to be emphasized for introduction. First, in
contrast to the ``natural'' \cite{Carl}, quickly and widely accepted
${\cal PT}-$symmetry assumption (\ref{pts}), the process of
acceptance of the second, formally equally important ${\cal
PCT}-$symmetry assumption (\ref{pcts}) appeared to be much slower
\cite{MZbook}. Moreover, it always looked, in the context of
phenomenology, too {\it ad hoc\,} and slightly suspicious (see,
e.g., a general theory in \cite{Geyer}, or a characteristic sample
of criticism in \cite{Lotor}). In this sense, our present approach
seems to offer a mathematically as well as phenomenologically very
persuasive new inner-product background for the introduction of the
charge ${\cal C}$. Secondly, our results will imply that the
standard version and interpretation of the role of the charge ${\cal
C}$ can be modified and generalized. In this sense, indeed, our
present letter may be read as an immediate and strong motivation for
the consideration of larger classes of quantum models using
non-Hermitian though not necessarily parity-times-time-reversal
symmetric Hamiltonians with real spectra.

\section{Quasi-Hermiticity}

The
assumption of the non-Hermiticity of
Hamiltonians seemed to be, initially, in a sharp
contradiction with Stone theorem
\cite{Stone},
i.e., with the well known correspondence
between the
unitarity of the evolution and the properties of the
underlying closed-system Hamiltonian.
Fortunately, the
puzzle
found an almost immediate resolution.
In essence (cf. review \cite{Carl}) it has been concluded
that
the Hamiltonians in question are
Hermitian
in a ``better-chosen'', amended Hilbert space ${\cal H}$.
The manifest non-Hermiticity of $H$
in $L^2(\mathbb{R})$
has been
declared inessential,
connected just with the auxiliary role of
the most common inner product
$\langle \psi_1|\psi_2\kt$ in $L^2(\mathbb{R})$.

For any ``false but favored'' Hilbert
spaces of the latter type we will
use, in what follows, the dedicated symbol ${\cal F}$.
One should add that even without
any reference to the above-mentioned antilinear symmetries
the simultaneous use of the two different inner products
(i.e., of the auxiliary, friendly one in ${\cal F}$  and of the
correct, more complicated but physical one in ${\cal H}$) was already a
part of an older modified version of quantum mechanics
called quasi-Hermitian
(see, e.g., its compact 1992 review \cite{Geyer}).
In this specific formulation of quantum theory
the ``correct Hilbert space''
${\cal H}$ appeared ``hidden''.
Represented, in  ${\cal F}$, via
an explicit formula
 \be
 \langle \psi_1|\psi_2\kt_{\cal H}=
 \langle \psi_1|\Theta|\psi_2\kt_{\cal F}\,
 \label{redede}
 \ee
which defines the correct physical
inner product in  ${\cal H}$ via its simpler partner in ${\cal F}$.

The correspondence (and non-equivalence) between Hilbert
spaces ${\cal F}$
and ${\cal H}$ is characterized by the
inner-product-metric operator
$\Theta$ which must be, i.a. \cite{Geyer}, positive definite and
self-adjoint,
 \be
 \Theta=\Theta^\dagger\ \ \ \ \ ({\rm in}\  {\cal F})\,.
 \label{onaro}
 \ee
As a consequence of the use of the two spaces ${\cal F}$
and ${\cal H}$
there is no problem with the coexistence
of the non-Hermiticity $H \neq H^\dagger$
in the auxiliary space ${\cal F}$
and the
``hidden'' Hermiticity of $H$ in ${\cal H}$.
In order to avoid confusion it is sufficient to
mark the latter property by another superscript, say, as follows,
 \be
 H=H^\sharp\ \ \ \ \ ({\rm in}\  {\cal H})\,.
 \label{zakl12}
 \ee
By construction, the latter, unitarity-guaranteeing
relation becomes equivalent to formula
 \be
 H^\dagger\,\Theta=\Theta\,H\ \ \ \ \ ({\rm in}\  {\cal F})
 \label{quass}
 \ee
i.e., to the $\Theta-$pseudo-Hermiticity {\it alias\,}
quasi-Hermiticity of $H$ in ${\cal F}$.
One can conclude that by relations~(\ref{redede}) and (\ref{quass}),
the picture of physics
is fully transferred from ${\cal H}$ to auxiliary ${\cal F}$.

\section{${\cal PT}-$ and ${\cal PCT}-$symmetries\label{kapjedna}}

The amendment and transfer of the older quasi-Hermitian
quantum mechanics of Ref.~\cite{Geyer}
to its
innovated version was
inspired by the 1998 letter \cite{BB}.
The essence of the success
of the innovation
can be seen in a simplification of the
technicalities due to the assumption that
the Hamiltonians of the form (\ref{BGr})
were special, viz.,
${\cal PT}-$ and ${\cal PCT}-$symmetric.
In this context, what was truly essential
was, first of all, the introduction of the charge ${\cal C}$.
This made the theory mathematically consistent
and, in the unbroken dynamical symmetry regime, fully
compatible with standard textbooks.
Simultaneously, the introduction of
the concept of charge
also helped to circumvent one of the main weaknesses
of the quasi-Hermitian quantum mechanics in which,
in the words of review \cite{Geyer},
the metric $\Theta=\Theta(H)$ ``is, in general,
not unique''.
For both of these reasons, the
growth of popularity of the
${\cal PT}-$ and ${\cal PCT}-$symmetric
quantum mechanics was truly impressive:
A concise outline of the whole story
can be found, e.g.,
in reviews \cite{ali,MZbook}.

New open questions also emerged:
we will mention some of them
in section \ref{quhe} below.
We will
emphasize there, once more, that
the success of the upgraded formalism
(called, usually, just  ${\cal PT}-$symmetric quantum mechanics)
was mainly given by the surprising
simplicity of its technical aspects.
The
main reason was that the
fathers-founders of ${\cal PT}-$symmetric quantum mechanics
\cite{BB}
complemented the fundamental requirement of the reality of
the spectrum
of $H$ by an apparently redundant ${\cal PT}-$symmetry (\ref{pts})
{\it alias\,}
parity-pseudo-Hermiticity \cite{ali} assumption
 \be
 H^\dagger\,{\cal P}={\cal P}\,H\,.
 \label{pseu}
 \ee
This was a fortunate decision.
The appeal and influence of such an assumption
(where ${\cal P}$ may but need not denote the operator of parity)
were not only
technical
(cf. the mathematically oriented reviews
\cite{ali,book})
but also intuitive and inspiring
(at present, the concept has applications
far beyond its original scope
\cite{Christodoulides,Carlbook}).
Indeed, relation (\ref{pseu})
can be re-read not only as the condition of self-adjointness
of $H$ in the Krein space endowed with the
indefinite (pseudo)metric
${\cal P}$ but also as
the property of an antilinear symmetry
$\,H\,{\cal PT}={\cal PT}\,H\,$ of $H$, with the
antilinear involution operator ${\cal T}$
carrying, as we already mentioned,
the physical meaning of time reversal \cite{BB,Carl,ali}.

In spite of such a rich phenomenological background
of the
${\cal PT}-$symmetry
{\it alias\,}
${\cal P}-$pseudo-Hermiticity  of $H$, by far the most
important (albeit not always emphasized) feature of the
${\cal PT}-$symmetric quantum theory of unitary systems
has to be seen in the role played by the other,
independent antilinear symmetry
$\,H\,{\cal PCT}={\cal PCT}\,H\,$ of $H$ were the operator
${\cal C}$ may but need not represent a charge \cite{ali}.
In any case, the product
${\cal PC}$
can be reinterpreted as
one of the most interesting
metric-operator solutions $\Theta=\Theta(H)$ of
Eq.~(\ref{quass}).
In other words,
the knowledge of the charge leads to the relation
 \be
 H^\dagger\,{\cal PC}={\cal PC}\,H\,
 \label{repseu}
 \ee
which guarantees the
unitarity of the evolution of the system
\cite{Carl,ali,Geyer,Lotor}.

The basic idea of our present note is that
in the area of physics using non-Hermitian
operators
the introduction of the two inner products (\ref{redede})
was in fact motivated mathematically.
One of the reasons lied in
the
complicated nature of
Hermitian conjugation of some Hamiltonians
(or of some other relevant operators $\Lambda$)
in ${\cal H}$.
Firstly, such a conjugation
[i.e., in the light of convention used in Eq.~(\ref{zakl12}),
the map $\Lambda \to \Lambda^\sharp$]
is inner-product dependent.
For this reason it makes sense to characterize it
by a subscripted antilinear operator ${\cal T}_{\cal H}$
which can easily be distinguished from its partner ${\cal T}_{\cal F}$
acting in ${\cal F}$.
Secondly,
after the
abbreviation of the action of ${\cal T}_{\cal H}$
or ${\cal T}_{\cal F}$ by the respective
dedicated superscripts $^\sharp$ and $^\dagger$,
the pull-down
of the conjugation from ${\cal H}$ to ${\cal F}$
[sampled by the replacement of Eq.~(\ref{zakl12}) by Eq.~(\ref{quass})]
acquired an explicit form,
 \be
 \Lambda^\sharp = \Theta^{-1}\,\Lambda^\dagger\,\Theta\,.
 \ee
In multiple applications the fully consistent
quasi-Hermitian model-building recipe
as described in review \cite{Geyer}
was successful. A simplification
of Eq.~(\ref{zakl12})
(representing the Hermiticity of $H$ in ${\cal H}$)
has been achieved
via its reduction to Eq.~(\ref{quass}) in ${\cal F}$.
Briefly, one can say that the initial
Hamiltonian-Hermiticity difficulty
was weakened.

\section{Another inner product and another quasi-Hermiticity\label{quhe}}

During the growth of popularity of the new paradigm it has been
revealed that the construction of the positive definite metric
$\Theta$ restricted  by the Hermiticity constraint (\ref{onaro}) may
be often the most difficult task in applications \cite{117}. The
techniques used in this setting may range from the direct
reconstruction of $\Theta=\Theta(H)$ using relation (\ref{quass}) up
to the sophisticated application of the vielbein formalism as
discussed in the very recent preprint \cite{2107}. A recommended
thorough review of these techniques is provided by chapter 4 of
review \cite{ali}. In principle, the desirable result may even be
just an indefinite pseudometric like ${\cal P}$ in Eq.~(\ref{pts})
(cf., e,g., formula number 78 in chapter~3 of review \cite{ali}).

The latter list of techniques can also be read as an inspiration and
starting point of our present considerations. Our basic idea is that
one can feel guided by the traditional quasi-Hermitian construction
pattern of Ref.~\cite{Geyer}. The replacement of a single inner
product (i.e., of a single Hilbert space ${\cal H}$) by the pair of
inner products (i.e., by a pair of non-equivalent Hilbert spaces
${\cal H}$ and ${\cal F}$) can be applied not only in the study of
Hamiltonian but also during the construction of the metric.

In the context of physics
the most promising aspect of the latter idea
is that
in the traditional ${\cal PT}-$symmetric
formalism as outlined in preceding section
one simplifies the mathematical construction
of the metric and, subsequently, the evaluation
of the probabilistic predictions
at an expense of narrowing the scope of the physical
model-building.
For this reason, any enhancement of flexibility of
the formalism is desirable.
One has to keep in mind that the conventional choice of the
Hamiltonian and charge forms just a very specific set of
the eligible quasi-Hermitian operators of observables
(see, e.g., \cite{arabky} or the
fairly general discussion of this point in \cite{Geyer}).
Also, in a way emphasized in \cite{ali}, even the traditional intuitive
${\cal PT}-$symmetry itself is a restriction
which may simplify some technicalities but which certainly narrows the
range of applicability of the non-Hermitian Hamiltonians with real spectra.

By Mostafazadeh \cite{ali} the generalized though
still positive-indefinite analogues of
parity
were denoted by the symbol $\eta$.
In proceedings \cite{pontr} we replaced $\eta$ by $Q$ and
used the operator in the role
of an inner-product metric in an
auxiliary
Pontryagin space.
Another form of simplicity of the model
has been achieved,
in Ref. \cite{positP}, with parity ${\cal P}$
replaced by its
positive definite
alternative ${\cal P}_+$.
The merits of the latter, extraordinary choice
(yielding in fact
another auxiliary Hilbert space)
were
illustrated by its relevance in
the study of an N-site-lattice Legendre-oscillator
toy-model Hamiltonian. Last but not least,
the latter choice of ${\cal P}_+$ has been shown, in \cite{scatt}, to
play a key role in a consistent
non-Hermitian (i.e., CST) description of scattering.

Whenever needed, the generic standard or non-standard realization
of the (possibly, generalized) parity will be denoted either by symbol
${\cal P}_g$ or, for simplicity, by ${\cal P}$.
The parallel
generalization of the charge
will be written, analogously, either as
${\cal C}_g$ or as
${\cal C}$.
Using this convention we are now prepared
to realize the
replacement
of a single inner product
by its two alternatives.
In other words,
the two older CST approaches based on the use of a
single Hilbert space ${\cal H}$,
or of a pair of
non-equivalent Hilbert spaces ${\cal H}$ and
${\cal F}$
will be
generalized via a
transition from the Hilbert-space doublet $[{\cal H},{\cal F}]$
to an inner-product-space triplet denoted as
$[{\cal H},{\cal R},{\cal F}]$.

For the sake of brevity, let us now
restrict attention just to the Hilbert-space
setup postponing, temporarily, the account of
the above-mentioned Krein-space alternative
to the next section.
This is a shortcut which
enables us to split the triplet of spaces into two
doublets.
In the first one, the
intermediate space ${\cal R}$
will play the role of a substitute for ${\cal F}$
with respect to ${\cal H}$.
In this case, we will
have to replace metric $\Theta$ in ${\cal F}$ by
the special positive charge ${\cal C}_+$ in ${\cal R}$.
In the second scenario,
${\cal R}$ will be a substitute for
${\cal H}$
with respect to ${\cal F}$.
Then we will replace $\Theta$ by ${\cal P}_+$.
Summarizing, we are now able to
upgrade Eq.~(\ref{redede}) as follows,
 \be
 \langle \psi_1|\psi_2\kt_{\cal H}=
 \langle \psi_1|{\cal C}_+|\psi_2\kt_{\cal R}\,,
 \ \ \ \ \ \
 \langle \psi_1|\psi_2\kt_{\cal R}=
 \langle \psi_1|{\cal P}_+|\psi_2\kt_{\cal F}\,.
 \label{ured}
 \ee
Due to our temporary assumptions, ${\cal C}_+$ does not represent a
charge, and also ${\cal P}_+$ is not parity.

In parallel, due care must be also paid to the transition from the
auxiliary antilinear operators $[{\cal T}_{\cal H},{\cal T}_{\cal
F}]$ to the triplet $[{\cal T}_{\cal H},{\cal T}_{\cal R},{\cal
T}_{\cal F}]$. By definition they play the role of the operators of
Hermitian conjugation in different Hilbert spaces so that in the
context of the ``upper'' two-space subset  $[{\cal H},{\cal R}]$,
the physical unitarity-guaranteeing hidden Hermiticity relation
(\ref{zakl12}) remains unchanged while, once we mark the action of
${\cal T}_{\cal R}$ by a new superscript $^\ddagger$,
Eqs.~(\ref{onaro}) and (\ref{quass}) become upgraded as follows,
 \be
 {\cal C}_+={\cal C}_+^\ddagger\ \ \ \ \ ({\rm in}\  {\cal R})\,,
 \label{uonaro}
 \ee
 \be
 H^\ddagger\,{\cal C}_+={\cal C}_+\,H\ \ \ \ \ ({\rm in}\  {\cal R})\,.
 \label{uquass}
 \ee
Similarly, in the context of
the ``lower'' two-space subset  $[{\cal R},{\cal F}]$, the
metric-Hermiticity
relation (\ref{uonaro}) in ${\cal R}$
becomes accompanied by its quasi-Hermiticity equivalent
in  ${\cal F}$,
 \be
 {\cal C}_+^\dagger\,{\cal P}_+={\cal P}_+\,{\cal C}_+\,.
 \label{ruseu}
 \ee
Concerning the Hamiltonian itself, a brief calculation
leads, finally, to the result
 \be
 H^\dagger\,\Theta=\Theta\,H\,,\ \ \ \ \ \Theta={\cal P}_+{\cal C}_+
 \label{upseu}
 \ee
which just reproduces the conventional
Eq.~(\ref{repseu}) above.

In order to avoid confusion it may also be useful to rewrite
Eq.~(\ref{ured}) in a more precise form. Thus, the conventional
product $\langle \psi_1|\psi_2\kt=\langle \psi_1|\psi_2\kt_{\cal F}$
(with the well known meaning, say, after the frequent choice of
${\cal F}=L^2(\mathbb{R})$) may be re-introduced via the definition
${\cal T}_{\cal F}: | \psi_1 \rangle \to \langle \psi_1|$ of the
ket-vectors in ${\cal F}$. In terms of this convention, the old
formula ${\cal T}_{\cal H}: | \psi_1 \rangle \to \langle \langle
\psi_1|$ (say, of Ref.~\cite{MZbook}) and the new, {\it ad
hoc}-notation formula ${\cal T}_{\cal R}: | \psi_1 \rangle \to
\langle \langle \langle \psi_1|$ define the more explicitly
characterized ket-vectors in ${\cal H}$ and in ${\cal R}$,
respectively. Ultimately, one can identify $\langle
\psi_1|\psi_2\kt_{\cal H} \equiv \langle \langle \psi_1|\psi_2\kt$
and write also $\langle \psi_1|\psi_2\kt_{\cal R}= \langle \langle
\langle \psi_1|\psi_2\kt$ in ${\cal F}$, therefore (note that these
conventions are different from those used, e.g., in \cite{tdthc}).

\begin{table}[h]
\caption{Hamiltonians $H$ in quantum theory using
one, two or three
Hilbert spaces.
}
 \label{dowe2} \vspace{.4cm}
\centering
\begin{tabular}{||c||c|c|c||c||c||}
    \hline \hline
     spaces & \multicolumn{3}{|c||}{operators}& comment\\
     \hline
       &
    \multicolumn{1}{|c|}{non-Hermitian}
    &
    \multicolumn{1}{|c||}{quasi-Hermitian}& {\rm Hermitian  }& \\
\hline
\hline
    ${\cal H}$
     & -  & - & $ H=H^\sharp\  $ & Ref.~\cite{Messiah}\\
    \hline \hline
    ${\cal H}$
     &
     -  & - &   $ H=H^\sharp\ $ & Ref.~\cite{Geyer}\\
     ${\cal F}$  &
     $ H\neq H^\dagger$ & $H^\dagger\, \Theta= \Theta\,H $&
      $ \Theta=\Theta^\dagger >0$ & $\Theta=$ metric \\
    \hline \hline
  ${\cal H}$
      &
     - & - &  $ H=H^\sharp\ $ & Eq.~(\ref{zakl12})\\
    ${\cal R}$  &
    $ H\neq H^\ddagger$ & $H^\ddagger\,{\cal C}={\cal C}\,H$
      & $ {\cal C}={\cal C}^\ddagger\ $  &${\cal C}=$ metric \\
     ${\cal F}$ &
     ${\cal C}
    \neq
    {\cal C}^\dagger$ & ${\cal C}^\dagger\,{\cal P}={\cal P}\,{\cal C}$
      & $ {\cal P}={\cal P}^\dagger$&${\cal P}=$ metric \\
      &
    $ H\neq H^\dagger$ & $H^\dagger\,\Theta=\Theta\,H$
    & $\Theta={\cal PC}=\Theta^\dagger$&$\Theta=$ metric \\
\hline \hline
\end{tabular}
\end{table}

Now we can return to the compact notation with ${\cal C}_+ \to {\cal
C}$ and with ${\cal P}_+ \to {\cal P}$. The situation is summarized
in Table~\ref{dowe2}. In a way guided by the quasi-Hermitian quantum
mechanics of Ref.~\cite{Geyer}, both of the positive inner-product
metrics ${\cal C}_+={\cal C}$ and ${\cal P}_+={\cal P}$ may be
treated as operators which are bounded, invertible and positive
definite, with bounded inverses. In such a scenario both of these
operators share the properties of their two-Hilbert-space
inner-product-metric predecessors. As a consequence, they also share
and extend their physical interpretation. A new, three-Hilbert-space
reformulation of the conventional unitary quantum mechanics is born.

\section{An alternative approach: Two auxiliary Krein spaces}

Strictly speaking, our latter conclusion is
temporary and formal but
still, it delivers an informal message which is nontrivial. It
can be formulated as follows.
After a removal of the requirement of positivity,
we can still work with the two independent general
inner-product metrics or pseudo-metrics ${\cal C}={\cal C}_g$
and ${\cal P}={\cal P}_g$.
The Table reflects
just an iterated application of the
two-inner-products trick. The physical message delivered
by the Table becomes different,
deserving a separate attention.
The point is that
besides the above-outlined three-Hilbert-space interpretation
of the relations summarized in Table \ref{dowe2},
they may be also given another, Krein-space-related,
interpretation.

An implicit emphasis upon
such an alternative
picture of quantum physics
motivated us to make the
indicative
choice of the notation with symbols
${\cal C}$ (for the first auxiliary inner-product metric or pseudo-metric in  ${\cal R}$)
and ${\cal P}$
(for the second auxiliary inner-product metric or pseudo-metric in  ${\cal F}$).
Such a notation
hinted at the possibility of treating
both of the spaces either as the
Hilbert spaces or
as the Krein spaces.
From the point of view of quantum theory
the latter, alternative point of view seems
equally interesting.
Due to a relaxation of the
metric-positivity constraints one can now
return to the narrower, more physics-oriented
(viz., charge and parity)
interpretations
of the respective operators.
In this way, the most traditional forms of the
parity and
Krein-space related
${\cal PT}-$symmetric quantum mechanics
as reviewed in \cite{Carl}
acquire the same formal
background and structure
as their
above-outlined three-Hilbert-space alternative.
The shared aspects of this correspondence
are summarized in Table~\ref{dowe}.


\begin{table}[h]
\caption{The Hilbert- and/or Krein-space interpretation of
${\cal PT}-$symmetric quantum mechanics.
}
 \label{dowe} \vspace{.4cm}
\centering
\begin{tabular}{||c||c|c|c|c||}
    \hline \hline
    quantum theory&
    \multicolumn{1}{|c|}{space}& operator
   & \multicolumn{1}{|c|}{\rm Hermitian conjugation   } & abbreviated\\
\hline
\hline
   Hermitian&  ${\cal H}$&   ${H}$
      &
      $H\,{\cal T}_{\cal H}={\cal T}_{\cal H}\,H\ $
        & $H=H^\sharp$ \\
    \hline
  quasi-Hermitian &   ${\cal H}$&   ${H}$
    &
      $H\,{\cal T}_{\cal H}={\cal T}_{\cal H}\,H\ $
        &  $H=H^\sharp$ \\ & ${\cal F}$&   ${\Theta}$
     &${\Theta}\,{\cal T}_{\cal F}={\cal T}_{\cal F}\,{\Theta}$
      & $\Theta=\Theta^\dagger$\\
    \hline
  ${\cal PT}-$symmetric & ${\cal H}$&   ${H}$
  &
       $H\,{\cal T}_{\cal H}={\cal T}_{\cal H}\,H\ $
      & $H=H^\sharp$ \\& ${\cal R}$&   ${\cal C}$
     &${\cal C}\,{\cal T}_{\cal R}={\cal T}_{\cal R}\,{\cal C}$
     & ${\cal C}={\cal C}^\ddagger$\\
    &  ${\cal F}$&   ${\cal P}$
    & ${\cal P}\,{\cal T}_{\cal F}={\cal T}_{\cal F}\,{\cal P}$
    &${\cal P}={\cal P}^\dagger$ \\
\hline \hline
\end{tabular}
\end{table}

One of the main
consequences of both of the latter two
interpretations of the
triple-inner-product-space formulations of quantum theory
is a complete reducibility
of the formalism
to its double-inner-product
predecessor.
A return
to the observable-quantity status of the
positive or indefinite
operators
${\cal P}$ and ${\cal C}$
renders it possible to
treat them simply as
preselected
dynamical-input factors of
the overall
physical Hilbert-space metric $\Theta={\cal PC}$ in ${\cal F}$.

On the purely pragmatic level
the sets of the operator relations listed in our two
Tables indicate that
we are free to eliminate the intermediate stage and to
skip not only the explicit use of the user-unfriendly
{\em physical\,}
Hilbert space ${\cal H}$ (as in the two-space formal regime)
but also the use of the less user-friendly
upper-auxiliary
Hilbert or Krein space
${\cal R}$.
The reason is that the
representation {\em of both of them\,} is now made
available, via positive definite $\Theta={\cal PC}$,
in the
single and preferable second auxiliary Hilbert space ${\cal F}$.

From the Hilbert/Krein-space
double-interpretation perspective let us add that
one of the not quite expected phenomenological consequences of our
formal results is that
even the generalized charge/metric ${\cal C}$ still represents
an observable quantity.
In the reformulated theory the proof is easy:
The observability property
${\cal C}^\dagger\,\Theta=\Theta\,{\cal C}$
of the charge (with respect to the physical metric $\Theta={\cal P C}$)
is just an immediate
formal consequence of Eq.~(\ref{ruseu}).

In another, last comment let us note that in both the Hilbert- and
Krein-space setups, another important consequence of the present
discovery of the correspondence between the introduction of the
auxiliary space ${\cal R}$ and of the auxiliary observable ${\cal
C}$ is that the theory now enables us to start building the models
in which many of the relevant operators could be chosen in a less
constrained, perceivably more elementary forms. An illustrative
example of such an option may be found outlined in the next section:
the choice of the illustration has been inspired by the multiple
formal difficulties encountered, in papers \cite{cpta} and
\cite{cptbe}, in supersymmetric setup.

\section{Differential operators\label{exa}}

For illustrative purposes let us consider the
most conventional non-Hermitian toy-model
Hamiltonian (\ref{BGr})
acting in
the auxiliary and user-friendly Hilbert space ${\cal F}=L^2(\mathbb{R})$.
For a direct, textbook-like treatment of this
Hamiltonian in the
correct and physical Hilbert space ${\cal H}$
(i.e., after Hermitization),
it would be necessary to construct the metric.
Unfortunately, the metric
would be represented
by a
hardly tractable, strongly non-local operator \cite{117}.
Moreover, whenever one decides to employ the mere single auxiliary
inner product, it would be comparably difficult to
find a sufficiently elementary
representation of the charge.
For both of these reasons it makes sense to
assume that
our Hamiltonian in question is ${\cal P}_g{\cal C}_g{\cal T}-$symmetric.

In the corresponding modified relation (\ref{pcts})
with the conventional
Hermitian conjugation ${\cal T}$, with the
generalized parity (i.e., with ${\cal P}$ replaced by ${\cal P}_g$), and
with the generalized charge (i.e., with ${\cal C}_g$ in place of
${\cal C}$) we may
work now with the less restricted families of operators
represented, say,
by the elementary differential expressions.
For the sake of brevity let us
choose, for example,
the generalized charge in its simplest first-order tentative form
 \be
 {\cal C}_g=\frac{d}{dx}+w(x)\,,
 \ \ \ \ \ w(x)=\sigma(x)+{\rm i}\alpha(x)
 \label{charge}
 \ee
with the two real functions
of a definite symmetry,
$\sigma(x)=\sigma(-x)$ and $\alpha(x)=-\alpha(-x)$.
In the Hamiltonian the potential
 \be
 V(x)=S(x)+L(x)+{\rm i}\Sigma(x)+{\rm i}\Lambda(x)\,
 \label{praseti}
 \ee
will be also split in such a way that
 \be
 S(x)=S(-x)\,,
 \ \ \ \
 L(x)=-L(-x)\,,\ \ \ \
 \Sigma(x)=\Sigma(-x)\,,\ \ \ \
 \Lambda(x)=-\Lambda(-x)\,.
 \ee
Under such an ansatz the antilinear symmetry relation
(\ref{pcts}) in its explicit pseudo-Hermiticity version (\ref{repseu})
can be treated as a compatibility constraint
and as an equation which defines
the admissible auxiliary metric ${\cal C}_g$
in terms of a given $H$, or vice versa.

The elementary form of both ${\cal C}_g$
and $H$ renders this illustrative problem solvable in closed form.
Indeed, it is entirely straightforward to show that
Eq.~(\ref{repseu}) degenerates, in such a case,
to the mere pair of relations
 \be
 S'(x)=2\,\sigma'(x)\sigma(x)-2\,\alpha'(x)\alpha(x)\,,
 \ \ \ \ \
 \Lambda'(x)=2\,\sigma'(x)\alpha(x)+2\,\alpha'(x)\sigma(x)\,
 \ee
where the primes denote the differentiation with respect to $x$.
This system is integrable and yields the closed-form solution
 \be
 S(x)= S(x,\omega)=\sigma^2(x))-\alpha^2(x)+\omega\,,
 \ \ \ \ \
 \Lambda(x)=2\,\sigma(x)\alpha(x)\,
  \label{prusiv}
 \ee
which contains just a single arbitrary integration constant $\omega$.
For an arbitrary input charge (\ref{charge})
the dynamical ${\cal PCT}-$symmetric
{\it alias\,}  ${\cal P}_g{\cal C}_g{\cal T}-$symmetric
quantum system will always exist
for Hamiltonians (\ref{BGr}) with potentials (\ref{praseti}) satisfying
the two explicit constraints (\ref{prusiv}) imposed upon the potential.

Due to the not too complicated form of relations (\ref{prusiv})
one could also change the formulation of the problem and treat the two
components $S(x)$ and $\Lambda(x)$ of the potential as an independent
dynamical input. Then, in a search for all of the eligible ``generalized charges''
${\cal C}_g$ it becomes sufficient to eliminate, say,
$\alpha(x)=\Lambda(x)/[=2\,\sigma(x)]$ and arrive at the
single implicit compatibility condition
 \be
 S(x)= \sigma^2(x)+\omega-\Lambda^2(x)/[4\sigma^2(x)]\,
 \ee
with the two different real
and non-vanishing eligible solutions $\sigma(x)$ such that
 \be
 2\,\sigma^2(x)=S(x)+\omega + \sqrt{[S(x)+\omega]^2+\Lambda^2(x)}\,.
 \ee
What now only remains to be verified is the
Hermiticity of the product ${\cal P}_g{\cal C}_g$
but this property can immediately be proved by direct insertion.
Incidentally, it appears to be equivalent to the
${\cal P}_g{\cal T}-$symmetry of ${\cal C}_g$. This
implies that
the generalized charge is also ${\cal P}_g{\cal C}_g{\cal T}-$symmetric, i.e.,
observable in the physical Hilbert space ${\cal H}$.

\section{Summary\label{summary}}

It has long been known that, in some sense,
the ${\cal PT}-$symmetric model-building recipe just
weakens the technical
difficulties
by their transfer from $H$
[controlled by Eq.~(\ref{zakl12})] to $\Theta$
[with the obligatory Hermiticity controlled by Eq.~(\ref{onaro})].
On this background
our present message can be summarized
as a recommendation of an application of the same
simplification strategy also
to the latter operator.

We have shown that a detailed realization of such a concept
is not entirely straightforward. Its core has been found to lie
in an introduction of a second auxiliary inner-product space
denoted by a dedicated symbol ${\cal R}$ and
lying, in some sense, in between the ``correct physical''
${\cal H}$
and the ``computation friendly''
${\cal F}$.
After an {\it ad hoc\,} amendment of the notation conventions,
the specific strength of the resulting
non-Hermitian model-building strategy
has been shown to lie in a weakening and replacement
of the Hermiticity of the metric
[cf. Eq.~(\ref{uonaro})]
by its quasi-Hermiticity or pseudo-Hermiticity
[sampled by Eq.~(\ref{ruseu})].

We pointed out that the
applicability of the new formalism
necessitates the reality of the spectrum of a given
diagonalizable candidate $H$ for the Hamiltonian.
At the same time, its users encounter an ambiguity
problem due to which
the set of the eligible Hermitizing
Hilbert-space metrics
[i.e., of the self-adjoint solutions $\Theta(H)$
of Eq.~(\ref{quass})]
is {\em too rich} in general.
In the conventional ${\cal PT}-$symmetric quantum mechanics
the uniqueness of
the physical metric $\Theta(H)={\cal PC}$ is being
based, therefore, on the choice of
a second {\it ad hoc\,} observable called charge.
Although this choice
can be perceived as somewhat arbitrary and
artificial (see also
its alternative as proposed in \cite{Lotor})
our present considerations
showed that
such a choice can in fact be given
a fairly deep mathematical meaning.

In this context, our observations and basic equations were
summarized
in two Tables.
Their contents emphasize that the
conventional
formulation of quantum mechanics in
Schr\"{o}dinger picture
(using just a single Hilbert space ${\cal H}$)
may be treated as formally equivalent
to the quasi-Hermitian formulation
(in which a pair of Hilbert spaces ${\cal H}$ and ${\cal F}$ is used)
as well as to the ${\cal PT}-$symmetric formulation.
Although the latter formulation is usually characterized
just by a specific choice of the metric $\Theta$,
the Tables offer an alternative, more satisfactory picture.
In it
one employs a
triplet of spaces $[{\cal H},{\cal R},{\cal F}]$
endowed with a triplet of different Hermitian conjugations.
In this approach, the symbols ${\cal C}$ and ${\cal P}$ may represent
either
the positive definite
Hilbert-space metrics [cf. Eq.~(\ref{ured})]
or the indefinite
Krein-space pseudometrics
(i.e., in the most popular special cases, the
charge and parity, respectively).
In both of these realizations of the idea, fortunately,
the formal outcome becomes almost independent of the
technical subtleties because
the pair of operators ${\cal P}$ and ${\cal C}$
only enters the condition of unitarity
(\ref{repseu})
in the form of their
positive definite and self-adjoint
product $\Theta={\cal PC}$.

Concerning the possible applicability of the present
metric-factorization idea in a broader quantum theoretical context
we have to admit that our attention has exclusively been paid here
to the mere unitary evolution studied in the hiddenly Hermitian
Schr\"{o}dinger picture. A decisive technical advantage of such a
restriction has been found in the fact that the underlying metric
$\Theta$ must necessarily remain stationary (or, more strictly
speaking, quasi-stationary, see the reasons given in \cite{ali}).
This made its factorization technically straightforward as well as
phenomenologically useful. We believe, nevertheless, that an
extension of the theory based on the present ``relegation of
Hermiticity'' could also successfully proceed in several other
directions.

In this sense we are particularly optimistic in the case of the
hiddenly Hermitian {\em Heisenberg\,} picture of Refs.~\cite{NHHP}.
We expect that the progress might be particularly quick there
because in this theory the metric necessarily remains
time-independent as well. In the second, technically more ambitious
direction of research based on the hiddenly Hermitian version of the
Dirac's interaction picture~\cite{tdth,tdthb} the situation remains
unclear at present. In the latter setting we would remain more
sceptical. The success seems to be an open question, indeed. Among
the most serious discouraging conceptual problems one finds that the
use of time-dependent metrics leads to the necessity of introduction
of the time-dependent ``Hamiltonians'' which {\em cease to be
observable}~\cite{ali,tdthc,tdthb,tdthf}. This might make all of the
constructive ``relegation of Hermiticity'' considerations much more
difficult \cite{2107,tdthbc}. Still, on positive side one also finds
important results like, e.g., the observation that the
time-dependence of the metrics $\Theta=\Theta(t)$ can be controlled
via suitable operator differential equations [see, e.g., equation
number 4 in \cite{tdthc}, or the examples of its solvability in
\cite{tdthb,tdthf}]. What is only missing here is the sharing of
notation conventions: Typically, the time-dependent ``Hamiltonians''
may be found denoted as $H_{gen}(t)$ (in \cite{tdth}) or as $G(t)$
(in reviews~\cite{tdthb,tdthbc}) or, in many papers, simply as
$H(t)$ (e.g., in \cite{tdthc,tdthf}), etc. Thus, what remains most
encouraging is only the current steady progress in the field.

\end{document}